\documentclass[aps,pre,reprint,superscriptaddress]{revtex4-2}
\usepackage{amsmath,graphicx,siunitx,xcolor,hyperref,bm,braket,physics2}
\usephysicsmodule{ab.legacy}
\hypersetup{colorlinks,linkcolor={blue!50!black},citecolor={blue!50!black},urlcolor={blue!80!black}}
\DeclareMathOperator*{\Var}{Var}

\begin{document}

\title{Characterising the slow dynamics of the swap Monte Carlo algorithm}

\author{Kumpei Shiraishi}

\affiliation{Laboratoire Charles Coulomb (L2C), Universit\'e de Montpellier, CNRS, 34095 Montpellier, France}

\author{Ludovic Berthier}

\affiliation{Laboratoire Charles Coulomb (L2C), Universit\'e de Montpellier, CNRS, 34095 Montpellier, France}

\affiliation{Gulliver, UMR CNRS 7083, ESPCI Paris, PSL Research University, 75005 Paris, France}

\date{\today}

\begin{abstract}
The swap Monte Carlo algorithm introduces non-physical dynamic rules to accelerate the exploration of the configuration space of supercooled liquids. Its success raises deep questions regarding the nature and physical origin of the slow dynamics of dense liquids, and how it is affected by swap moves. We provide a detailed analysis of the slow dynamics generated by the swap Monte Carlo algorithm at very low temperatures in two glass-forming models. We find that the slowing down of the swap dynamics is qualitatively distinct from its local Monte Carlo counterpart, with considerably suppressed dynamic heterogeneity both at single-particle and collective levels. Our results suggest that local kinetic constraints are drastically reduced by swap moves, leading to nearly Gaussian and diffusive dynamics and weakly growing dynamic correlation lengthscales. The comparison between static and dynamic fluctuations shows that swap Monte Carlo is a nearly optimal local equilibrium algorithm, suggesting that further progress should necessarily involve collective or driven algorithms. 
\end{abstract}

\maketitle

\section{Introduction}

When glass-forming liquids are cooled, molecular motion becomes very slow while their structure remains disordered; they eventually become glasses~\cite{Ediger_1996}. The glass transition is an important research subject that has been studied over many years using both experiments and theory~\cite{Ediger_1996,Berthier_Biroli_2011}. Numerical studies are vitally important because simulations can provide complete information about microscopic particle motion near the glass transition~\cite{BerthierReichman2023}. However, computational studies of the glass transition are inherently difficult. Close to the transition, slow molecular motion results in rapidly growing relaxation times, and equilibrium sampling of the configuration space becomes difficult within reasonable CPU times. At low temperatures, conventional molecular dynamics (MD)~\cite{Frenkel_Smit} or local Monte Carlo (MC)~\cite{Allen_Tildesley,Berthier_2007} methods become inefficient. To achieve sampling near the experimental glass transition, various enhanced techniques have been implemented~\cite{BarratBerthier2023}. For instance, the parallel tempering method allows a faster exploration of the rugged free energy landscape~\cite{Hukushima_1996,Yamamoto_2000,Jung2024Normalizing}. Instead, event-chain and lifted Monte Carlo algorithms propose collective particle moves to improve the local Metropolis Monte Carlo approach~\cite{Bernard_2009,isobe2016applicability,PhysRevLett.133.028202,berthier2024monte}. Recently, machine learning techniques were also applied to sample low-temperature configurations~\cite{ciarella2023machine,Jung2024Normalizing}. Despite good progress, none of these methods was able to provide substantial improvement over conventional MD and MC techniques so far.

For certain models of supercooled liquids, the swap Monte Carlo (SMC) method~\cite{kranendonk1991computer,Grigera_2001} completely changes the situation~\cite{Berthier_2016,Ninarello_2017}. In recent years, the SMC method has attracted significant attention, leading to the development of several continuous or discrete polydisperse particle models optimized for this method~\cite{Ninarello_2017,Berthier_2D_2019,Fullerton_2020,Parmar_ultrastable_2020,Jung2023}. In such models, SMC is remarkably efficient, accelerating relaxation processes by more than ten orders of magnitude compared to local MC and MD methods. Therefore, it has become possible to access very low temperature regimes that were previously inaccessible. Various aspects of glass physics were then investigated, such as thermodynamics~\cite{Berthier_2017,Berthier_2D_2019,Guiselin_SM_2022}, plasticity~\cite{Ozawa_2018,Richard_plasticity_2020}, thermal vibrations~\cite{Wang_low_freq_2019}, and two-level systems~\cite{Khomenko_2020,Mocanu2023}. Additionally, this technique allows for in-silico generation of particle configurations with stability comparable to the one achieved experimentally by physical vapor deposition~\cite{Swallen_2007,Ediger_2017}, thus offering useful insights into the physics of ultrastable glasses~\cite{Berthier_Charbonneau_Flenner_Zamponi_2017,Berthier_2020,Herrero2023PNAS,Herrero2023JCP}.

Thanks to SMC, new insights were also gathered regarding the low-temperature relaxation processes in supercooled liquids~\cite{Guiselin_Excess_2022}. It has long been known that the relaxation of liquids near the glass transition is spatially heterogeneous~\cite{Ediger_2000,Donati_PRL_1999,Berthier_DH_2011,Keys_2013}. By allowing numerical studies at much lower temperatures and longer times, SMC was used to clarify the relative role of rare activation events and dynamic facilitation~\cite{Chandler_2010} in the formation of dynamic heterogeneity in model supercooled liquids~\cite{Guiselin_Excess_2022,Scalliet2022Thirty,Herrero2024PRL,chacko2024dynamical}.

Surprisingly, no such detailed real-space characterization of relaxation processes is available for the dynamics produced by SMC itself, which also becomes slow at low temperatures. This gap is presumably explained by the non-physical nature of the particle swaps introduced by SMC, which makes the Monte Carlo dynamics somewhat artificial. However, the several explanations proposed to account for the success of SMC all make assumptions and predictions about the influence of particle swaps on the relaxation dynamics which have not been tested directly. In particular, the role of local kinetic constraints~\cite{Wyart_2017,Gavazzoni2024Testing,Gutirrez2019}, dynamic facilitation and activated processes~\cite{Berthier_2019}, the existence of a shifted mode-coupling crossover temperature~\cite{Ikeda_Zamponi_Ikeda_2017,Szamel_2018,Brito_2018} were invoked. It would be desirable to connect these theoretical discussions with detailed real-space characterisation of the relaxation processes generated by the SMC dynamics. Such analysis could also provide useful guidance in the context of recent attempts to improve SMC by incorporating event-driven and collective swap moves~\cite{PhysRevLett.133.028202,berthier2024monte}. 

Here, we characterize the slow dynamics and its spatial and temporal heterogeneity in SMC dynamics by performing simulations of two glass-forming models in which SMC provides a very large acceleration of the dynamics. We characterise the dynamic heterogeneity at the single-particle and collective levels, emphasizing differences and similarities with the physical dynamics obtained using conventional MC. We find that the dynamic heterogeneity in SMC dynamics is much weaker and the relaxation is spatially more homogeneous. Interestingly, the dynamic length scale controlling the SMC dynamics is very close to the static point-to-set length scale, which governs thermodynamic fluctuations associated to the approach to a random first-order transition~\cite{Kirkpatric_1989,Bouchaud_2004,Montanari_2006}. This suggests that for the studied models, SMC is a nearly optimal local equilibrium algorithm.  

The manuscript is organised as follows. 
In Sec.~\ref{sec:model} we present the two models and numerical methods used.
In Sec.~\ref{sec:bulk_relaxation}, the bulk relaxation is characterized for normal MC and SMC. Then, we characterize the dynamic heterogeneity of the relaxation.
In Sec.~\ref{sec:single_DH}, we present indicators of dynamic heterogeneity at the single-particle level, while Sec.~\ref{sec:collective} characterizes the collective behaviour of the relaxation and its relation to static quantities. 
In Sec.~\ref{sec:correlation_MD}, we quantify the correlation between MC and SMC dynamics. 
Finally, we conclude in Sec.~\ref{sec:discussion} where we present a theoretical discussion of the numerical results and their implications. 

\section{Models and numerical Methods}

\label{sec:model}

We consider systems of continuously polydisperse particles~\cite{Ninarello_2017,Berthier_2D_2019} enclosed in a square ($D=2$) or cubic box ($D=3$) with periodic boundary conditions. Two particles $i$ and $j$ separated by a distance $r_{ij}$ interact via a repulsive potential
\begin{align}
v(r_{ij}) = \epsilon \left(\frac{\sigma_{ij}}{r_{ij}}\right)^{12} + c_0 + c_2 \left(\frac{\sigma_{ij}}{r_{ij}}\right)^2 + c_4 \left(\frac{\sigma_{ij}}{r_{ij}}\right)^4,
\end{align}
if $r_{ij} < 1.25 \sigma_{ij}$; the potential is zero otherwise. The constants $c_0$, $c_2$, and $c_4$ are chosen to ensure the continuity of the potential up to its second derivative at the cutoff. A non-additive interaction rule $\sigma_{ij} = 0.5(\sigma_i + \sigma_j)(1 - 0.2\abs{\sigma_i - \sigma_j})$ is employed to avoid crystallization and phase separation at low temperatures~\cite{Ninarello_2017}. The diameters $\{ \sigma_i \}$ of the particles are drawn from a distribution $P(\sigma) \propto 1/\sigma^3$ in a finite range $[0.73,1.62]$. The number of particles is $N$ and the linear length $L$ of the simulation box is determined from the number density $\rho = N/L^D = 1.0$ ($D = 2, 3$) using the average particle diameter as unit length. In both two- (2D) and three-dimensions (3D), we consider systems with $N=1000$. The temperature is expressed in units of $\epsilon$ taking the Boltzmann constant to unity. 

At each temperature $T$, the system is first equilibrated by the SMC method. For each MC attempt, we randomly pick up one particle $i$, located at $\bm{r}_i$, and move its position to $\bm{r}_i + \delta\bm{r}$~\cite{Berthier_2007}, with $\delta\bm{r}$ a random vector in a cube of linear length $\delta = 0.15$. In addition to translational motion, a swap move exchanging the diameters of two randomly chosen particles is also proposed with a probability of 0.2. Both types of MC proposed moves are accepted according to the Metropolis condition. We define $N$ attempted moves as one MC step, which defines our time unit. After equilibrium configurations are obtained by SMC, we perform either SMC or standard MC simulations where no swap move occurs.

\section{Bulk relaxation dynamics}

\label{sec:bulk_relaxation}

\subsection{Self-intermediate scattering function}

We start by presenting a comparison of the bulk dynamics between MC and SMC. To this end, we compute the self-intermediate scattering function
\begin{align}
F_s(q,t) = \Braket{\frac{1}{N}\sum_{i=1}^N \cos(\bm{q}\cdot\Delta\bm{r}_i(t))},
\label{eq:fsqt}
\end{align}
where $\Delta\bm{r}_i(t) = \bm{r}_i(t) - \bm{r}_i(0)$, and the brackets indicate an ensemble average. The motion of the center of mass of the system is removed. The wavevectors are chosen as $q = 6.7$ in 3D~\cite{Scalliet2022Thirty}, $q = 6.5$ in 2D~\cite{Berthier_2D_2019}. In the 2D system, we use the cage-relative displacements $\Delta \bm{r}_i^\mathrm{CR}(t)$ (removing the motion of the center of mass of the neighbors of particle $i$) instead of $\Delta\bm{r}_i(t)$, to minimize the influence of Mermin-Wagner fluctuations~\cite{Mazoyer_2009,Illing_2017}.

\begin{figure}
\includegraphics[width=\linewidth]{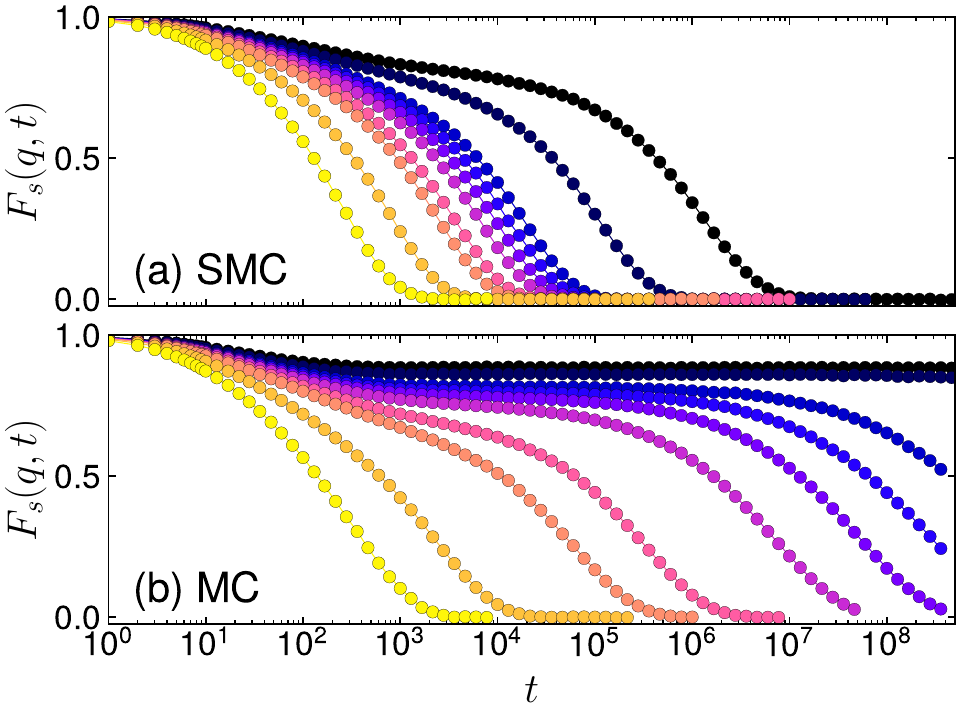}
\caption{The self-intermediate scattering function $F_s(q,t)$ defined in Eq~(\ref{eq:fsqt})  for (a) SMC and (b) MC dynamics in 3D. Each color corresponds to a temperature, with $T=0.400$, 0.200, 0.130, 0.117, 0.102, 0.095, 0.090, 0.085, 0.070, and 0.060, from left to right in both panels. A similar two-step decay emerges at low $T$ in both dynamics, but SMC relaxes much faster than MC at any given $T$.}
\label{fig:fsqt}
\end{figure}

In Fig.~\ref{fig:fsqt}, $F_s(q,t)$ is presented for both MC and SMC dynamics for temperatures ranging from $T=0.400$ to $T=0.060$ in the 3D system. At high temperature, $T=0.400$, which is above the onset temperature $T_o \approx 0.266$ in 3D~\cite{Ozawa_2019}, $F_s(q,t)$ decays with a single relaxation step in both SMC and MC dynamics. As $T$ is lowered, a plateau develops in $F_s(q,t)$ in the MC dynamics, and the relaxation of $F_s(q,t)$ then proceeds with the well-known two-step relaxation process~\cite{Berthier_Biroli_2011}. At temperatures around and below the mode-coupling crossover temperature, $T_\mathrm{MCT} = 0.107$~\cite{Ninarello_2017}, it is no longer possible to observe the entire decay of $F_s(q,t)$ within our simulation time window when using the MC dynamics. By contrast, the relaxation of SMC does not show the same slowing down observed in normal MC dynamics when lowering temperatures. The emergence of a plateau with SMC occurs at a much lower temperature, near $T \approx 0.080$, but its complete decay can easily be observed even at temperatures much below $T_\mathrm{MCT}$. These results agree with previous results~\cite{Ninarello_2017,Berthier_swap_MD_2019}, and show that the dynamic relaxation is very much accelerated by the introduction of swap moves. Similar data are obtained in the 2D system (not shown). Overall, the SMC results show that a form of slow dynamics, characterised by a two-step decay of bulk correlation function is also emerging at sufficiently low temperatures for SMC. In the remainder, we characterise this slow dynamics in detail.  

\subsection{Structural relaxation time}

\begin{figure}
\centering
\includegraphics[width=\linewidth]{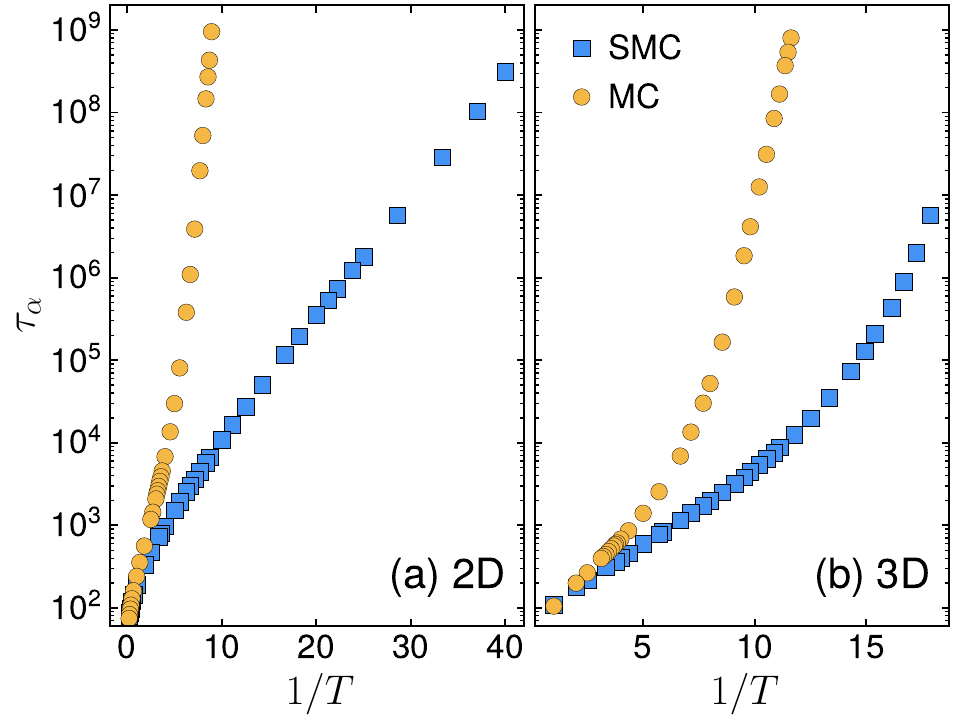}
\caption{Temperature dependence of the structural relaxation time $\tau_\alpha$ of SMC and MC dynamics in (a) 2D and (b) 3D.}
\label{fig:tau_alpha}
\end{figure}

We characterize the acceleration offered by the swap moves by measuring the structural relaxation time $\tau_\alpha$, defined as $F_s(q,t=\tau_\alpha) = e^{-1}$. As shown in Fig.~\ref{fig:tau_alpha}, the structural relaxation time of SMC is much shorter than the one of MC at each temperature. The first conclusion, known from previous work on SMC~\cite{Ninarello_2017,Berthier_2D_2019}, is that SMC can reach thermal equilibrium at temperatures much lower than conventional MC.

Let us take a closer look at the temperature evolution of the structural relaxation times in Fig.~\ref{fig:tau_alpha}. In 2D, the SMC data shows a nearly Arrhenius behaviour down to very low temperatures, with a temperature scale which is much smaller than for MC dynamics. This leads to a very large acceleration of the relaxation dynamics~\cite{Berthier_2D_2019}. 

At high temperatures in 3D, the SMC dynamics shows an Arrhenius growth of $\tau_\alpha$ down to an onset temperature $T \approx 0.1$, which is much lower than the onset temperature for MC dynamics, $T \approx 0.25$. We can also compare the location of the mode-coupling crossovers for both dynamics by fitting the relaxation times to a power-law, $\tau_\alpha \sim (T - T_\mathrm{MCT})^{-\gamma}$, as predicted by the mode-coupling theory of the glass transition~\cite{Gotze_2008}. We obtain $T_\mathrm{MCT} = 0.144$ in 2D and $T_\mathrm{MCT} = 0.107$ in 3D for MC dynamics, consistent with previous estimates~\cite{Ninarello_2017,Scalliet2022Thirty}. For the SMC dynamics, we find $T_\mathrm{MCT} = 0.036$ in 2D and $T_\mathrm{MCT} = 0.055$ in 3D. The quality of the fits in all cases is similar, and the power describes about 3 decades of the evolution of the relaxation times. These fits are consistent with a shift of the location of the mode-coupling transition by the swap moves~\cite{Ikeda_Zamponi_Ikeda_2017,Szamel_2018,Brito_2018,kuchler2023understanding}, but given the modest quality of the fits, it is difficult to draw firm conclusions on this point.  

\section{Single-particle dynamic heterogeneity}

\label{sec:single_DH}

\subsection{Van Hove distributions}

A major feature of dynamic relaxation processes near the glass transition is the presence of dynamic heterogeneity~\cite{Ediger_2000,Donati_PRL_1999,Berthier_DH_2011}. This name generically refers to the existence of spontaneous fluctuations of the local dynamics, with the coexistence of fast and slow regions in the material. While this is well-documented for supercooled liquids relaxing via physical dynamics such as MC~\cite{Berthier_2007,Berthier_JCP_I_2007} and MD~\cite{Kob_DH_1997,Chaudhuri_2007}, we wish to analyse the behaviour obtained from the SMC dynamics. 

We start by analyzing the van Hove distribution function defined as 
\begin{align}
P(\Delta r(t)) = \Braket{\frac{1}{DN} \sum_{i} \sum_\alpha \delta \left[
\Delta r - ( r_{i\alpha}(t) - r_{i\alpha}(0) )
\right]
},
\end{align}
where $\delta(x)$ is the Dirac delta function and $\alpha = x$, $y$ (and $z$ for $D=3$). Notice that since diameters fluctuate in time in the SMC dynamics, it makes little sense to distinguish particles by their sizes to record their dynamics. 

\begin{figure}
\includegraphics[width=\linewidth]{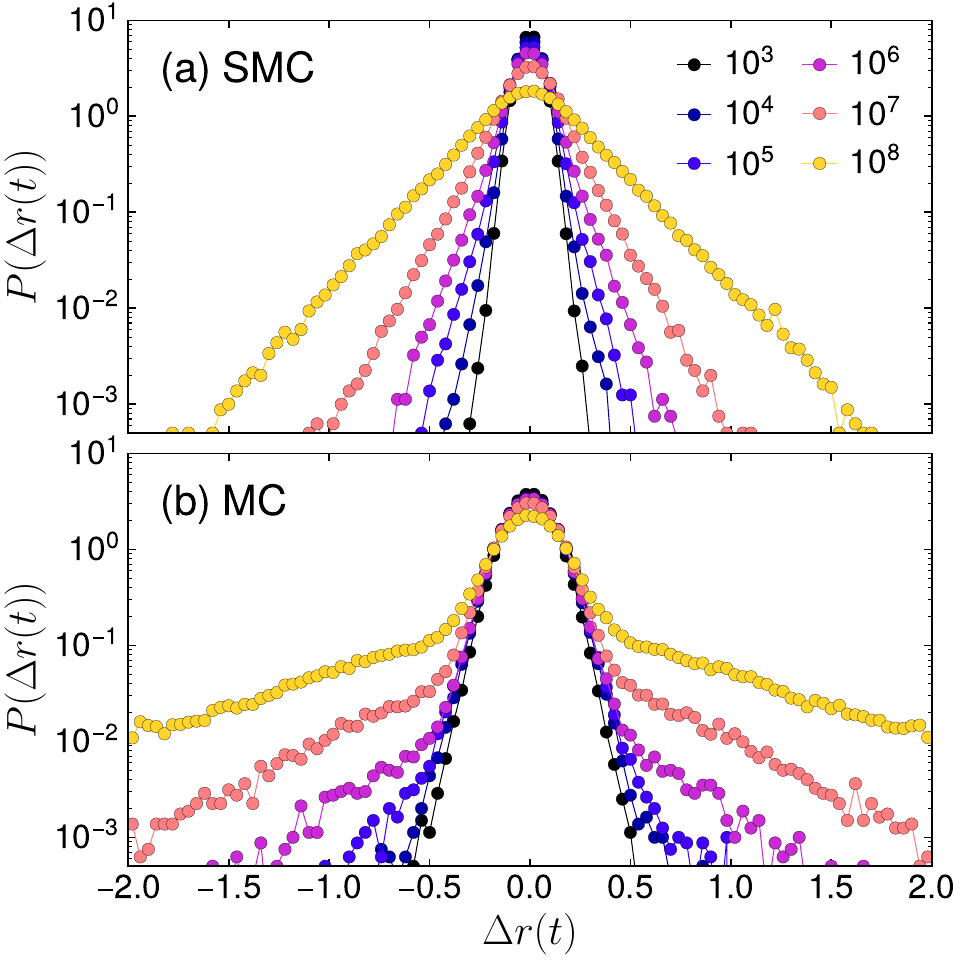}
\caption{The van Hove distribution function at different times in 2D for (a) SMC at $T=0.025$ where $\tau_\alpha= \num{3.2e8}$ and (b) MC at $T=0.115$ where $\tau_\alpha=\num{4.3e8}$. The non-Gaussian nature of particle distribution observed in MC dynamics is considerably suppressed for SMC.}
\label{fig:vanHove2D}
\end{figure}

\begin{figure}
\includegraphics[width=\linewidth]{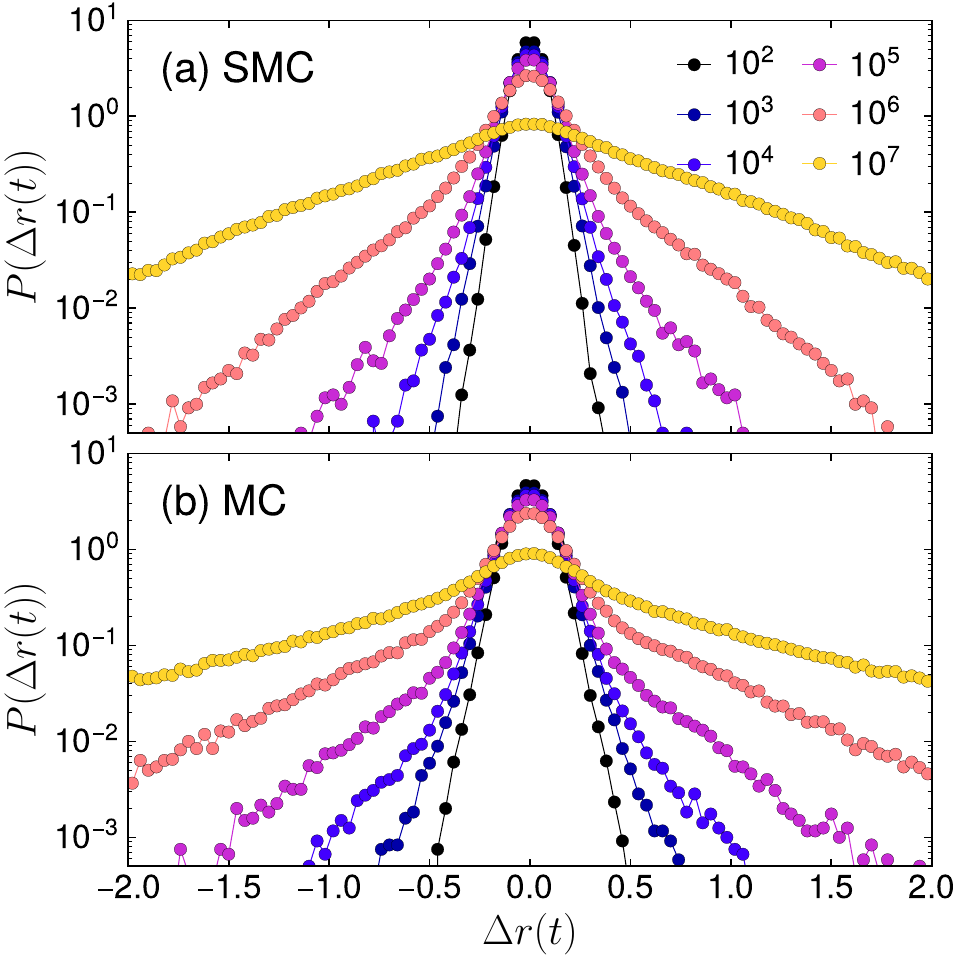}
\caption{The van Hove distribution function at different times in 3D for (a) SMC at $T=0.058$ where $\tau_\alpha= \num{2.0e6}$ and (b) MC at $T=0.105$ where $\tau_\alpha=\num{1.8e6}$. The non-Gaussian nature of particle distribution observed in MC dynamics is considerably suppressed for SMC.}
\label{fig:vanHove3D}
\end{figure}

In Figs.~\ref{fig:vanHove2D} and \ref{fig:vanHove3D}, the time dependence of the van Hove function in MC and SMC is shown for 2D and 3D systems, respectively. In these figures, temperatures are chosen so that structural relaxation times $\tau_\alpha$ are approximately the same for MC and SMC to make the comparison meaningful. 

For MC dynamics, $P(\Delta r(t))$ has a Gaussian shape at short times, as expected from thermal motion within a solid. However, as time increases, the distribution develops non-Gaussian tails that signal the presence of particles which relax much faster than the bulk and start exploring large distances at times much shorter than $\tau_\alpha$~\cite{Kob_DH_1997,Chaudhuri_2007}. At intermediates times, the van Hove distribution appears clearly bimodal, and is well described at least qualitatively, as the superposition of two populations of particles. Such property of single-particle displacements is well-known and universal among many glass-formers when dynamics is slow~\cite{Chaudhuri_2007}. The van Hove distribution returns to a Gaussian distribution in the limit of large times only, $t/\tau_\alpha \to \infty$~\cite{berthier2023comment}. 

The data for the SMC dynamics, by contrast, show much weaker signs of non-Gaussianity. The distributions are again close to a Gaussian at short times, as observed in MC. However, as time increases, the distributions seem to broaden in a relatively homogeneous manner, with no clear emergence of broad exponential tails. A clear separation between mobile and immobile particles is not observed in SMC dynamics even as time increases towards $\tau_\alpha$. This observation suggests that the dynamic heterogeneity is suppressed in SMC dynamics. The comparison between 2D and 3D data also shows that the suppression is somewhat weaker in 3D than in 2D, which correlates with the relative efficiency of the SMC equilibration speedup observed in the two models.  

\subsection{Non-Gaussian parameter}

To quantitatively characterize the non-Gaussianity of the distribution, we calculate the non-Gaussian parameter $\alpha_2(t)$~\cite{Rahman1964}, which is generally expressed as
\begin{align}
\alpha_2 (t) = \frac{D}{D+2} \frac{\braket{r^4(t)}}{\braket{r^2(t)}^2} - 1,
\end{align}
in $D$ dimensions~\cite{Biroli2022}, so that $\alpha_2=0$ if the van Hove distribution is Gaussian.

\begin{figure}
\includegraphics[width=\linewidth]{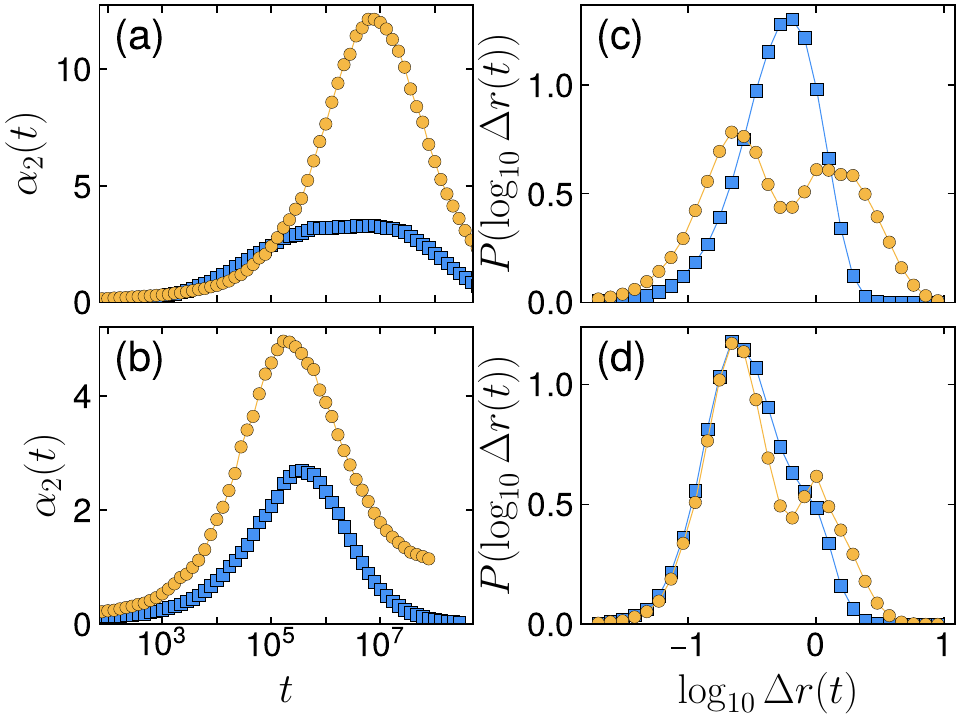}
\caption{The non-Gaussian parameter $\alpha_2(t)$ for SMC (blue squares) and MC (yellow circles) dynamics in (a) 2D and (b) 3D. In 2D, SMC data are for $T=0.025$ ($\tau_\alpha = \num{3.2e8}$) and MC data for $T=0.115$ ($\tau_\alpha = \num{4.3e8}$). In 3D, SMC data are for $T=0.058$ ($\tau_\alpha = \num{2.0e6}$) and MC data for $T=0.105$ ($\tau_\alpha = \num{1.8e6}$).
The distributions $P(\log_{10}\Delta r(t))$ in (c) 2D for $t=\num{4.6e8}$ and (d) 3D for $t=\num{1.7e6}$.}
\label{fig:alpha2}
\end{figure}

In Fig.~\ref{fig:alpha2}(a), $\alpha_2(t)$ of SMC and MC are shown for the 2D system, for temperatures chosen to have comparable relaxation times. Both functions have the expected non-monotonic time dependence with a peak at a time about 10 times shorter than $\tau_\alpha$, indicating maximal deviation from Gaussianity. However, the peak height is considerably lower for SMC than for MC (by a factor of about 3) in agreement with the distributions shown in Fig.~\ref{fig:vanHove2D}. A similar suppression is observed in the 3D system in Fig.~\ref{fig:alpha2}(b), but the suppression is a bit smaller (by a factor of about 2). 

As a complementary view, we present the probability distribution of the logarithm of particle displacements $P(\log_{10}\Delta r(t))$ for times near the peak of $\alpha_2(t)$. In this representation~\cite{Flenner2005}, the MC dynamics displays a bimodal shape with two distinct peaks, while the SMC distributions do not. These results confirm that the drastic reduction of the non-Gaussian parameter indeed results from a weaker distinction between fast and slow particles in the relaxation processes emerging from the SMC dynamics. 

\begin{figure}
\includegraphics[width=\linewidth]{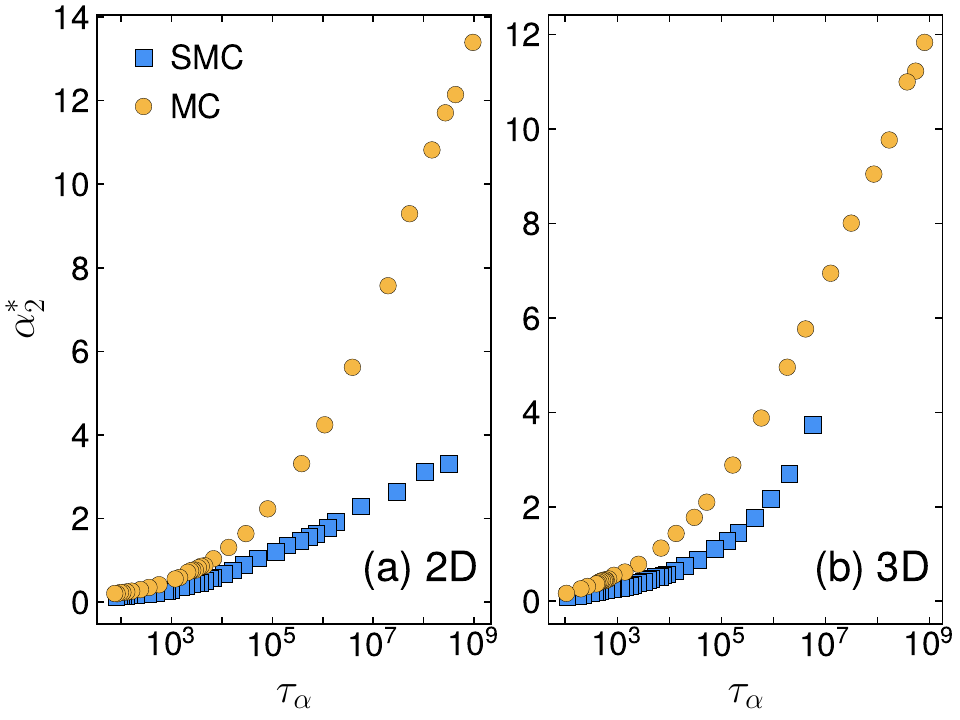}
\caption{The temperature evolution of the peak value $\alpha_2^*$ of the non-Gaussian parameter for SMC and MC dynamics in (a) 2D and (b) 3D is represented using a parametric plot against the relaxation time $\tau_\alpha(T)$. The non-Gaussian parameter is strongly suppressed by the SMC dynamics, even when the two dynamics are compared at equal relaxation times, with a more spectacular suppression in 2D.}
\label{fig:alpha2star}
\end{figure}

Finally, we follow the temperature evolution of the peak of $\alpha_2(t)$, which we denote $\alpha_2^*$, see Fig.~\ref{fig:alpha2star}. As observed in Fig.~\ref{fig:tau_alpha}, both dynamics display very different relaxation times at the same temperature $T$. Therefore, it makes sense to compare them at equal values of the relaxation time $\tau_\alpha$, as we adopt in Fig.~\ref{fig:alpha2star} where the temperature is used as a parametric variable. Even in this representation, the suppression of $\alpha_2^*$ in SMC dynamics remains obvious. In 3D, $\alpha_2^*$ starts to increase at very low temperatures for SMC, while in 2D there is no such tendency and $\alpha_2^*$ remains very modest down to extremely low temperatures. 

Overall, we conclude from this section that the single-particle dynamics in SMC exhibits a very modest dynamic heterogeneity with a very weak contrast between fast and slow particles, leading to nearly Gaussian van Hove distributions, no broad exponential tails, and small values of the non-Gaussian parameter compared to the conventional MC dynamics. 

\subsection{Decoupling between self-diffusion and structural relaxation}

An important manifestation of the presence of dynamic heterogeneity in supercooled liquids is the well-documented decoupling between the self-diffusion coefficient $D_s(T)$ and the bulk viscosity $\eta(T)$~\cite{Ediger_2000}. While both quantities are related by the Stokes-Einstein relation in simple liquids, $D_s \sim T/\eta$, this relation is violated when both $D_s$ and $\eta$ vary by orders of magnitude upon approaching the glass transition temperature. The resulting decoupling phenomenon has attracted significant attention in experimental~\cite{Ediger_2000,Cicerone_1996,Swallen_2003} and numerical and theoretical~\cite{Tarjus_1995,Berthier_2004,kumar2006nature,Eaves2009,PhysRevLett.119.056001,Kawasaki_2017,Das2022} studies.

Here, we study whether decoupling is observed in SMC dynamics, and compare the results with MC dynamics. To analyse the decoupling, it is more convenient to compare the temperature evolution of the self-diffusion constant defined from the mean-squared displacements, $D_s = \lim_{t \to \infty} \braket{\abs{\Delta \bm{r}_i(t)}^2}/(2Dt)$ to the one of the structural relaxation time $\tau_\alpha(T)$. It is known that $\tau_\alpha$ and $\eta$ have typically a similar temperature dependence~\cite{Kawasaki_2017}. For simple liquids obeying simple diffusive motion, the structural relaxation time $\tau_\alpha(q,T)$ measured at a given wavevector $q$ and temperature $T$ is related to $D_s(T)$ as $\tau_\alpha(q,T) \sim 1/(D_s(T) q^2)$. In this Fickian limit, the product $D_s \times \tau_\alpha$ is a constant independent of temperature~\cite{Berthier2005EPL}. We study deviations from this Fickian reference for the two models at hand.     

\begin{figure}
\centering
\includegraphics[width=\linewidth]{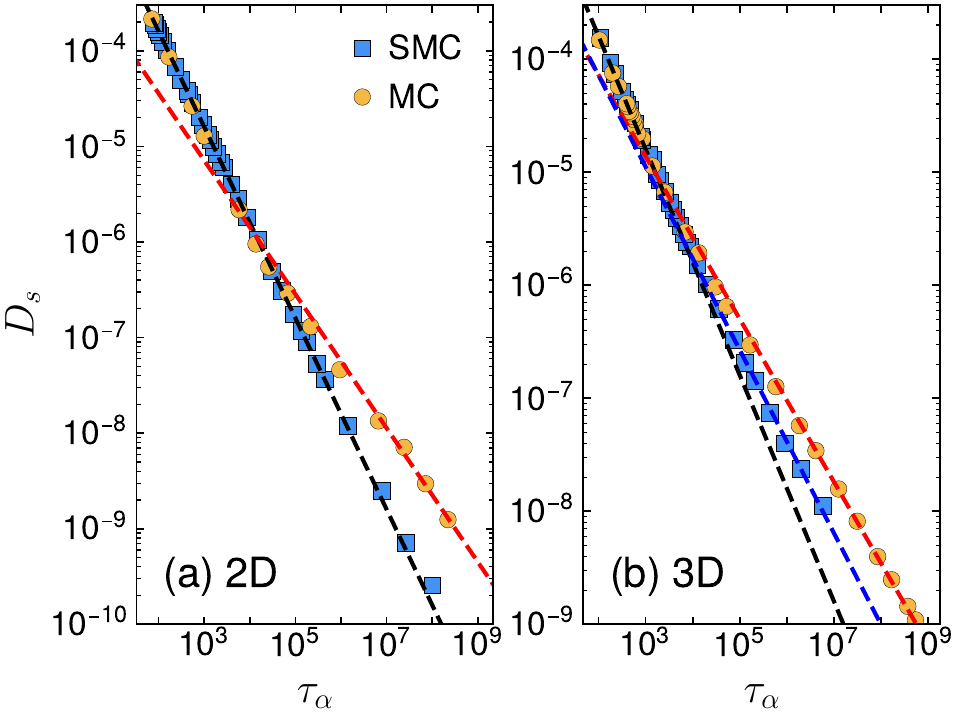}
\caption{The structural relaxation time dependence of the self diffusion constant $D_s$.
The black dashed line in both panels shows $D_s \sim \tau_\alpha^{-1}$.
The red dashed line shows $D_s \sim \tau_\alpha^{-0.70}$ in the 2D panel $D_s \sim \tau_\alpha^{-0.72}$ in the 3D panel.
The blue dashed line in the 3D panel shows $D_s \sim \tau_\alpha^{-0.81}$.}
\label{fig:decoupling}
\end{figure}

The analysis of decoupling in Fig.~\ref{fig:decoupling} shows the parametric evolution of $D_s$ with $\tau_\alpha$ as temperature is varied. In MC dynamics, decoupling occurs and the relation $D_s \sim \tau_\alpha^{-1}$ breaks down as temperature is lowered. A fractional relation describes the decoupling accurately, $D_s \sim \tau_\alpha^{-\zeta}$ with values $\zeta = 0.70$ and $\zeta = 0.72$ observed in 2D and 3D, respectively, instead of the expected $\zeta=1$ in the absence of decoupling. In agreement with the suppression of single-particle heterogeneity, we also find that decoupling is strongly suppressed for SMC dynamics in both 2D and 3D, with exponents $\zeta = 0.81$ in 3D and $\zeta = 1$ in 2D. Therefore, the suppression of dynamic heterogeneity is accompanied by a drastic suppression of decoupling, which is vanishingly small in 2D. We have confirmed these conclusions by analysing the wavevector dependence of the structural relaxation time $\tau_\alpha$~\cite{Berthier_2004,Berthier2005EPL} and found indeed that diffusive dynamics is very nearly recovered for 2D while deviations from diffusive behaviour are very small in 3D.  

Finally, we briefly discuss the influence of Mermin-Wagner fluctuations on decoupling in 2D. In Fig.~\ref{fig:decoupling}, we used normal displacements $\Delta \bm{r}_i(t)$ instead of cage-relative displacements $\Delta \bm{r}_i^\mathrm{CR}(t)$ to compute both $D_s$ and $\tau_\alpha$. Employing cage-relative displacements removes the influence of Mermin-Wagner fluctuations on $\tau_\alpha$ for relatively large wavevectors, but this correction leads to an incorrect estimate of the diffusion constant $D_s$~\cite{PhysRevLett.123.265501,Li2019}. To avoid this artefact, we decided to use normal displacements in the $D_s$ calculations and we chose a relatively modest system size ($N=1000$) to reduce the influence of Mermin-Wagner fluctuations on $\tau_\alpha$. We have checked that a similar amount of decoupling is observed when $\tau_\alpha$ is calculated from the cage-relative definition of $F_s(q,t)$, so that our conclusions regarding the comparison between MC and SMC are robust. 

\section{Collective dynamics}

\label{sec:collective}

\subsection{Four-point dynamic susceptibility}

The analysis of single-particle dynamics revealed that the distinction between the fastest and slowest particles is suppressed by SMC dynamics. Yet, these fluctuations do not completely vanish and in any trajectory spontaneous dynamic fluctuations remain present. In this section we ask whether these fluctuations are spatially correlated. For MC dynamics, a large number of studies have clarified the presence and temperature evolution of growing dynamic lengthscales~\cite{Berthier_DH_2011,Karmakar_2014}.  We now perform a similar analysis for the SMC dynamics. 

\begin{figure}
\includegraphics[width=\linewidth]{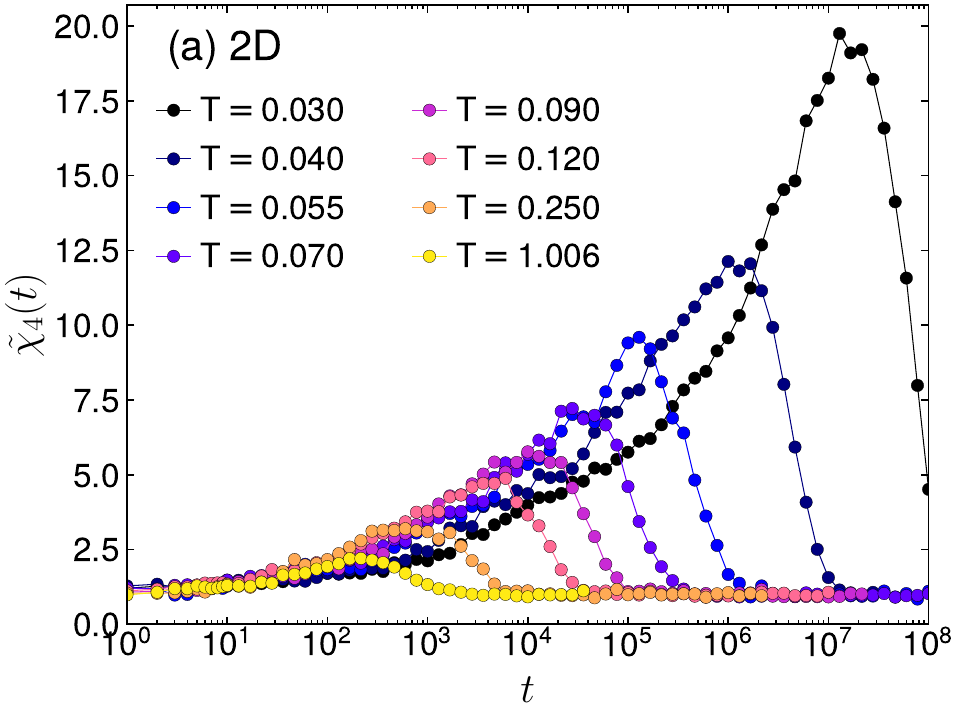}
\includegraphics[width=\linewidth]{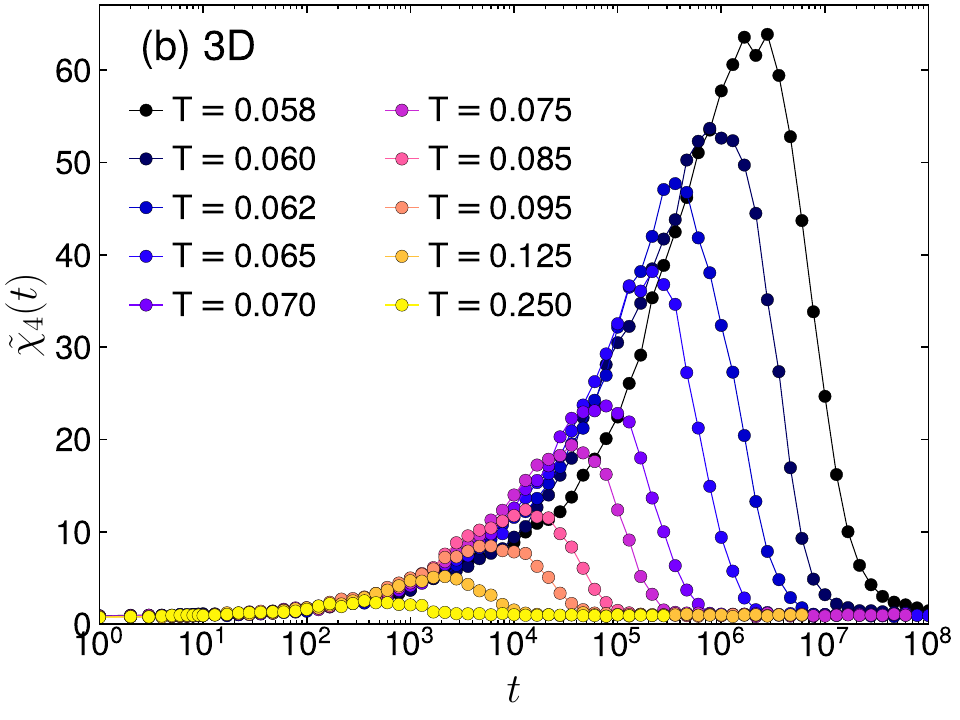}
\caption{Time dependence of the normalised dynamic susceptibility $\tilde{\chi}_4(t)$ for the SMC dynamics in (a) 2D and (b) 3D at different temperatures.}
\label{fig:chi4}
\end{figure}

To this end, we compute the four-point dynamic susceptibility, which is the standard tool for measuring dynamic heterogeneity~\cite{Franz_1999,Glotzer2000,lacevic2003spatially,Toninelli_2005}. We define the dynamic susceptibility from the spontaneous fluctuations of the instantaneous value of the self-intermediate scattering function $\hat{F}_s(q,t)$ around its average $F_s(q,t)$ in Eq.~(\ref{eq:fsqt}): 
\begin{align}
\chi_4(t) = N \left[
\Braket{\hat{F}_s(q,t)^2} - \Braket{\hat{F}_s(q,t)}^2
\right].
\end{align}
By definition, we have $F_s(q,t) = \Braket{\hat{F}_s(q,t)}$. To normalise $\chi_4(t)$, it is convenient to consider the situation where no correlation exists between particles. In that case, $\chi_4(t)$ reduces to the single-particle quantity $\chi_4^s(t) = \Var(f_s^i)$, with $f_s^i = \cos({\bm q} \cdot \Delta {\bm r}_i)$ the single-particle contribution to $F_s(q,t)$, and $\Var(x) = \braket{x^2} - \braket{x}^2$. We then define a rescaled four-point function 
\begin{align}
\tilde{\chi}_4(t) = \frac{\chi_4(t)}{\chi^s_4(t)}.
\label{eq:chiresc}
\end{align}
As defined in Eq.~(\ref{eq:chiresc}), $\tilde{\chi}_4(t) \approx 1$ at both very short and very long times, when interparticle correlations play no role, while the amplitude of $\tilde{\chi}_4(t)$ at intermediate roughly reflects the volume of dynamic correlations~\cite{Toninelli_2005,Berthier_JCP_I_2007}.

The normalised dynamic susceptibility $\tilde{\chi}_4(t)$ for SMC dynamics in 2D and 3D is shown in Fig.~\ref{fig:chi4} for a broad range of temperatures. As expected, it is close to unity at both short and long times, and it develops a peak at a timescale which is slaved to the average structural relaxation time $\tau_\alpha$, as frequently observed for conventional MD dynamics~\cite{Glotzer2000,Karmakar_2009,Coslovich_EPJE_2018,Das2022,Berthier_2007,Berthier_JCP_I_2007}. Importantly, the amplitude of the peak of the susceptibility increases far above unity as temperature decreases and dynamics slows down. 

\begin{figure}
\includegraphics[width=\linewidth]{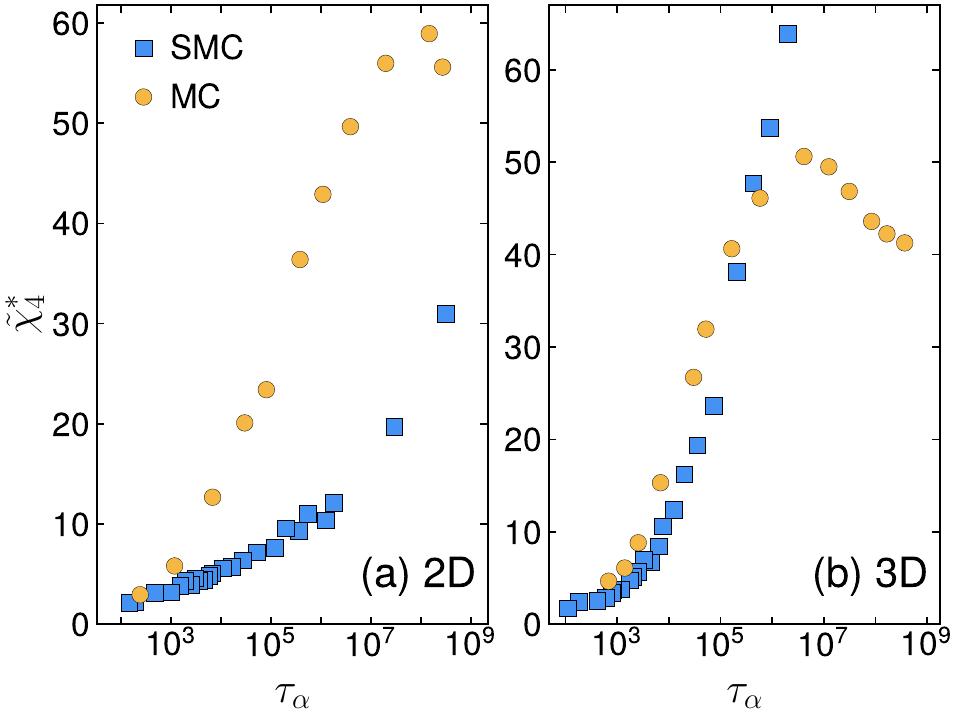}
\caption{Parametric evolution of the peak height of the dynamic susceptibility $\tilde{\chi}^*_4(T)$ as a function of the structural relaxation time $\tau_\alpha(T)$ for both MC and SMC dynamics in (a) 2D and (b) 3D.}
\label{fig:chi4_star}
\end{figure}

The peak height of the dynamic susceptibility represents the typical volume of spatially correlated domains~\cite{Toninelli_2005}. We denote it as $\tilde{\chi}_4^*(T)$  and present its evolution with temperature in Fig.~\ref{fig:chi4_star}, where we again use a parametric representation against the average relaxation time $\tau_\alpha(T)$ to compare MC and SMC dynamics on an equal footing. 

Starting with 3D and the MC dynamics, we observe the usual behaviour where $\tilde{\chi}_4^*$ grows by lowering the temperature, but exhibits a much slower growth below the mode-coupling crossover temperature near $T_\mathrm{MCT} \approx 0.107$~\cite{Ninarello_2017}. A similar crossover behaviour has been widely reported before in other models of glass-forming liquids~\cite{Coslovich_EPJE_2018,Das2022,Ortlieb2023}. In the SMC dynamics, $\tilde{\chi}_4^*$ evolves very similarly when represented against the relaxation time, see Fig.~\ref{fig:chi4_star}(b), but we observe no sign of a plateau or a saturation at the lowest studied temperature, perhaps we cannot reach as long timescales with SMC compared to MC dynamics. At low temperatures, the roughly linear relation between $\tilde{\chi}_4^*$ and $\log \tau_\alpha$ suggests that the relaxation time grows exponentially fast with the correlation volume. We note that these data can also be described with a power law $\tilde{\chi}_4^* \sim \tau_\alpha^{0.2}$, with a fairly small exponent that is often compatible with a logarithmic relation.  

The comparison is again different in the 2D system. First, in the normal MC dynamics, no clear crossover to the plateau is observed, perhaps because the temperature range we observed ($T \geq 0.117$) is not much below $T_\mathrm{MCT} = 0.120$. A more important difference is the comparison between MC and SMC dynamics. While in 3D both dynamic susceptibilities grow similarly with $\tau_\alpha$, in the 2D system, the dynamic susceptibility is much smaller both when compared at the same temperature and when compared at the same value of the relaxation time. At the longest relaxation time we could access, the dynamic susceptibility for SMC is about twice smaller than the one in MC. 

\subsection{Dynamic and static correlation lengthscales}

\begin{figure}
\includegraphics[width=\linewidth]{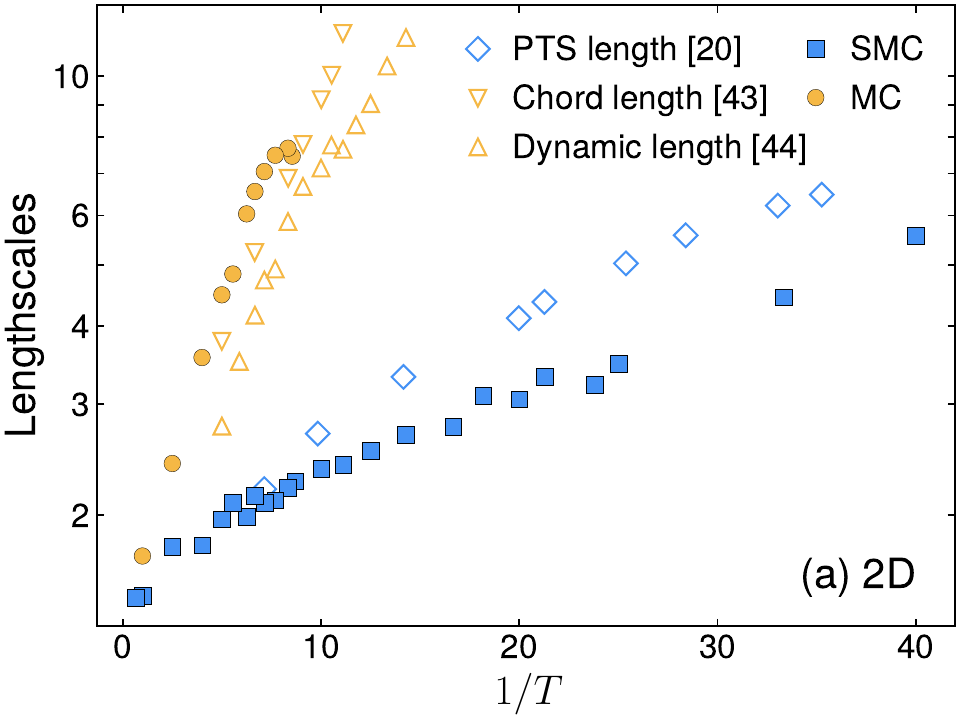}
\includegraphics[width=\linewidth]{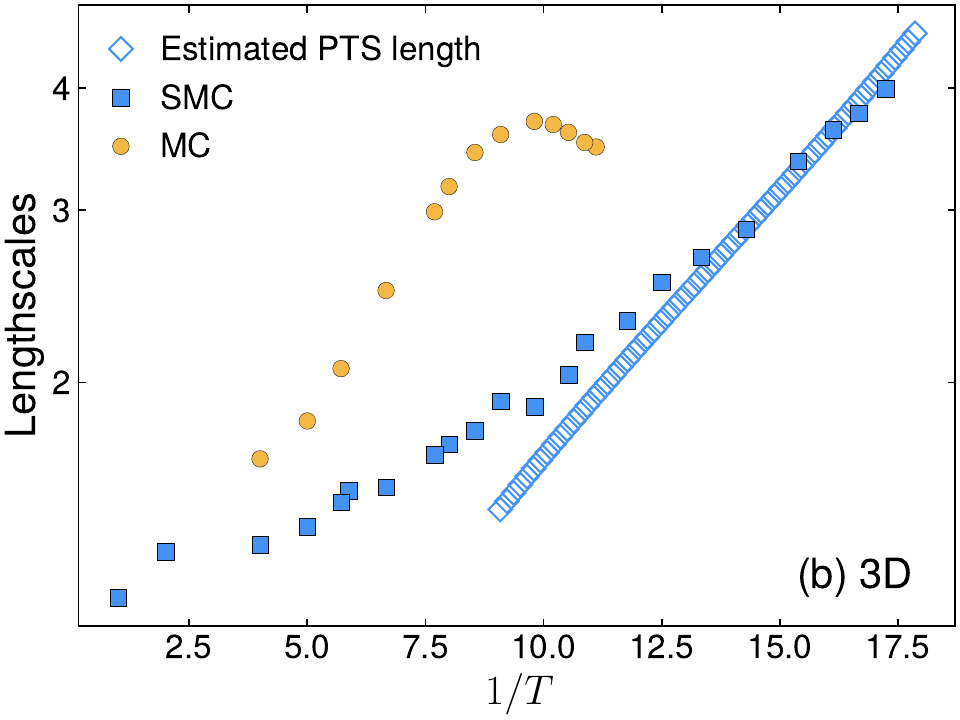}
\caption{Temperature evolution of various estimates for static and dynamic lengthscales in (a) 2D and (b) 3D, using an Arrhenius representation.}
\label{fig:lengths}
\end{figure}

The observation that dynamic susceptibilities have distinct values in MC and SMC dynamics at the same temperature is important as this directly suggests that the peak value $\tilde{\chi}_4^*$ is sensitive to the microscopic dynamic rules and is therefore not uniquely controlled by static, thermodynamic or structural fluctuations which are, by definition, insensitive to the chosen dynamics. 

Here we discuss this point in more depth and collect a number of estimates for correlation lengthscales discussed before for the present models in order to understand better their interrelation, if any. To this end, we first convert the measured dynamic susceptibitities into correlation lengthscales using the proxy 
\begin{equation}
    \xi_d = \left( \tilde{\chi}_4^* \right)^{1/D}, 
    \label{eq:xid}
\end{equation}
thus giving us estimates for the dynamic correlation lengthscale $\xi_d$ for both dynamics in both 2D and 3D. The results are shown in Fig.~\ref{fig:lengths}, which represents the temperature evolution of the resulting dynamic lengthscales for both 2D and 3D. 

In 2D, we can compare the resulting dynamic lengthscale to two independent estimates obtained via completely different methods: the average chord length extracted in Ref.~\cite{Scalliet2022Thirty} and the dynamic lengthscale obtained from the inhomogeneous geometries studied in Ref.~\cite{Herrero2024PRL} for the same 2D system. The data shown in Fig.~\ref{fig:lengths}(a) show a good degree of agreement for the MC dynamics and provide a consistent trend for the temperature evolution of the dynamic lengthscale characterising the MC dynamics.  

When we add the dynamic lengthscale estimated for the SMC dynamics using Eq.~(\ref{eq:xid}) to the same figure we confirm that the amplitude and temperature dependence of this SMC dynamic correlation lengthscales is totally different than the three dynamic lengthscales obtained from MC dynamics. The SMC dynamic correlations are much shorter-ranged, and grow slower with decreasing the temperature. This decoupling between the two types of correlation lengthscales indicate that they have a different physical origin and in particular cannot be both the direct result of some underlying thermodynamic or structural correlations.  

A decoupling between dynamic and static correlation lengthscales has been reported before in computer simulations~\cite{Charbonneau_decorrelation_2013,Scalliet2022Thirty}. For the present 2D model in particular, a comparison between the static point-to-set lengthscale~\cite{Montanari_2006,Bouchaud_2004,Biroli_2008} and the average chord length confirmed this suggestion~\cite{Scalliet2022Thirty}. Therefore, we extract the point-to-set length from the literature for our 2D system~\cite{Berthier_2D_2019} and compare it to the dynamic lengthscales in Fig.~\ref{fig:lengths}(a). This comparison confirms the observation that dynamic lengthscales from the physical MC dynamics are decoupled from the static point-to-set lengthscale. Interestingly, our results in Fig.~\ref{fig:lengths}(a) indicate that the growth of the dynamic lengthscale characterising the SMC dynamics resembles very closely the evolution of the point-to-set lengthscale.

We now perform a similar analysis for the 3D system in Fig.~\ref{fig:lengths}(b), but we have less literature results to guide our analysis. Firstly, no alternative measurement of the dynamic correlation lengthscale is available for the 3D system studied here. However, our own data are sufficient to conclude, as for the 2D system that the SMC dynamic correlation lengthscale is smaller and grows more slowly than the one characterising the MC dynamics. Regarding static correlations, the point-to-set correlation lengthscale has not been measured for our 3D system of polydisperse soft spheres. However, the configurational entropy $S_{\rm conf}(T)$ has been estimated~\cite{Ozawa_JCP_2018}, which can be used to infer the point-to-set lengthscale as $(TS_\mathrm{conf})^{-2/3}$~\cite{Kirkpatric_1989,Berthier_2017}. We add this estimate in Fig.~\ref{fig:lengths}(b), using an arbitrary vertical shift to convert this number into a physical lengthscale with a behaviour compatible with known results for the point-to-set lengthscale for similar systems~\cite{Berthier_2017}. Despite these caveats, we again find that for the 3D system the point-to-set lengthscale is strongly decoupled from the dynamic correlation lengthscale for the MC dynamics, but is more strongly correlated with the one characterising the SMC dynamics. 

\subsection{Adam-Gibbs relation}

\begin{figure}
\includegraphics[width=\linewidth]{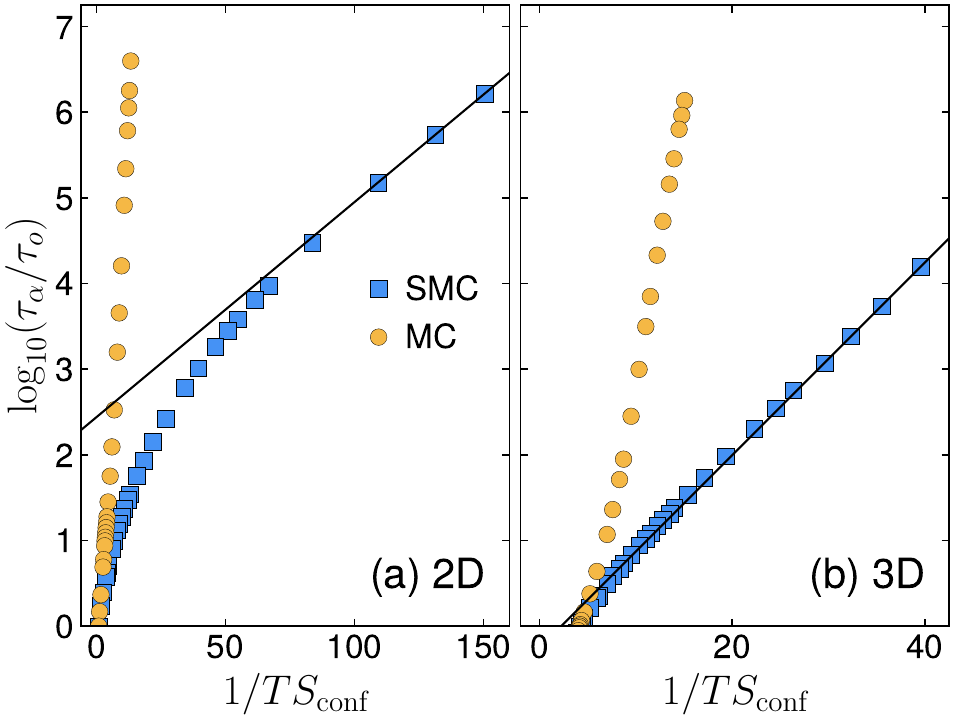}
\caption{Adam--Gibbs plots for SMC and MC dynamics in (a) 2D and (b) 3D. Black lines represent the Adam--Gibbs relation~\eqref{eq:AG} fitted from SMC data at low temperatures.}
\label{fig:AG}
\end{figure}

The observation that MC and SMC dynamic correlation lengthscales are respectively decoupled and coupled to numerical estimates of static correlation lengthscales encourages us to revisit the Adam-Gibbs relation~\cite{Adam_Gibbs_1965} that connects the configurational entropy to the bulk relaxation time. This relation was first formulated by Adam and Gibbs in order to infer a dynamic slowing down from a decreasing configurational entropy, which had been noticed by Kauzmann~\cite{Kauzmann_1948}. The formula was then revisited and incorporated in the more modern framework of the random first-order transition (RFOT) theory~\cite{simple_glasses}, and in particular in its phenomenological extension to finite dimensions~\cite{Kirkpatric_1989,Bouchaud_2004}.   

In this view, the Adam-Gibbs relation is the mathematical formulation of the physical statement that the sharp decrease of the configurational entropy $S_{\rm conf}(T)$ is causally responsible for the rapid growth of the bulk relaxation time $\tau_\alpha(T)$ as the temperature decreases towards the glass transition:
\begin{align}
\log \left( \frac{\tau_\alpha}{\tau_o} \right) \propto \frac{1}{TS_\mathrm{conf}}, 
\label{eq:AG}
\end{align}
where $\tau_o$ is a microscopic timescale, roughly corresponding to the value of the structural relaxation time at the onset temperature. Several numerical~\cite{Sastry_2001,PhysRevLett.109.095705,handle2018adam} and experimental~\cite{Ediger_1996,Richert_Angell_1998} studies have tested the validity of this relation. 

We investigate if and how this relation is obeyed for the two types of dynamics studied here for the same models. By construction, the right-hand side of Eq.~(\ref{eq:AG}) is a purely thermodynamical quantity, while the left-hand side depends on the microscopic dynamics. A test of the Adam-Gibbs relation amounts to a parametric plot of the relation between $\tau_\alpha$ and $S_{\rm conf}$, using temperature as a running parameter, see Fig.~\ref{fig:AG}. For the quantity $\tau_o$, we use the value of $\tau_\alpha$ of our dynamics data at the onset temperatures $T_o$ as determined in Ref.~\cite{Ozawa_2019} ($T_o = 1.006$ in 2D and $T_o = 0.266$ in 3D). For $S_\mathrm{conf}$, we use the fits to the data obtained in Ref.~\cite{Ozawa_2019}. 

Over the limited time window allowed by the simulations, we find an apparent linear relation between $\log(\tau_\alpha)$ and $1/TS_{\rm conf}$, in good agreement with Eq.~(\ref{eq:AG}). This was noticed before for the present models in the context of MC dynamics~\cite{Ozawa_2019} as well as in many other models~\cite{Sastry2000}. Interestingly, we find that the SMC dynamics also obeys the Adam-Gibbs relation in the low-temperature regime. Together with the observation that the SMC dynamic correlation lengthscale is comparable to the point-to-set lengthscale (while the MC one is not), this suggests that the SMC dynamics in fact obeys more closely the physical picture predicted by RFOT theory and captured by the Adam-Gibbs relation.  

Pushing this line of thought to its limit, we can use the results in Ref.~\cite{Ozawa_2019} regarding the relation between entropy and point-to-set lengthscale to infer numerical values for the dynamic exponent $\psi$ that expresses the relation between $\tau_\alpha$ and the point-to-set lengthscale. For the SMC dynamics, we obtain by this reasoning the value $\psi \approx 0.9$ in 2D and $\psi \approx 1.2 - 1.7$ for 3D. Although these estimates are of course limited by the modest time window accessible to the simulations, they are fully compatible with the predicted range $\psi \in [d/2, d]$, whereas the MC results are usually less satisfactory and lead to quite small values of $\psi$, especially in 3D where $\psi = 0.75-0.9$ was found for MC dynamics~\cite{Ozawa_2019}. (See Ref.~\cite{Capaccioli_2008} for an experimental counterpart to the latter finding.)

As a final note, we remark that testing the validity of the Adam-Gibbs relation by a parametric plot of $\log(\tau_\alpha)$ against $S_{\rm conf}$ is not easy as the data often appear strongly correlated and are nearly linearly related. This is illustrated in Fig.~\ref{fig:AG} where the two dynamics appear to satisfy the Adam-Gibbs relation reasonably well, which seems contradictory. A much stronger test of the connection between statics and dynamics is the strong correlation of dynamic and static lengthscales, as in Fig.~\ref{fig:lengths}, which more directly probes the underlying physics. This issue is discussed further in the final section.  

\section{Are swap and physical dynamics correlated?}

\label{sec:correlation_MD}

\subsection{Maps of isoconfigurational relaxation}

So far, the comparison between MC and SMC dynamics established many differences: the relaxation times are different, the heterogeneity of the single-particle dynamics is different, spatial correlations of the dynamics and links to thermodynamic fluctuations also differ. 

These results may appear surprising given the long line of research that has tried to establish deep connections between structure and dynamics in supercooled liquids~\cite{Kawasaki_2007,Royall_2015,hallett2018local,Tanaka_2019}. How can the same structure be a good predictor of two widely different dynamics? To address this question, it is useful to raise an intermediate point and ask how strongly MC and SMC dynamics are correlated with one another. 

We directly compared maps of local relaxation dynamics produced by MC and SMC, and their relation to the underlying structure of the supercooled liquid. To do so, we use the isoconfigurational ensemble~\cite{Widmer_Cooper_2004} and run a large number of trajectories using either MC or SMC starting from the same initial condition. In practice, we first prepare equilibrium configurations using the SMC algorithm. Starting from each configuration, we perform both MC and SMC simulations. Trajectories are generated by using different sets of random numbers. In both cases, the number of isoconfigurational trajectories is 400. We characterize the local relaxation by the self-intermediate scattering function measured for each particle $f^i_s(q,t) = \cos \left( \bm{q}\cdot\Delta\bm{r}_i^\mathrm{CR}(t) \right)$.

\begin{figure}
\includegraphics[width=\linewidth]{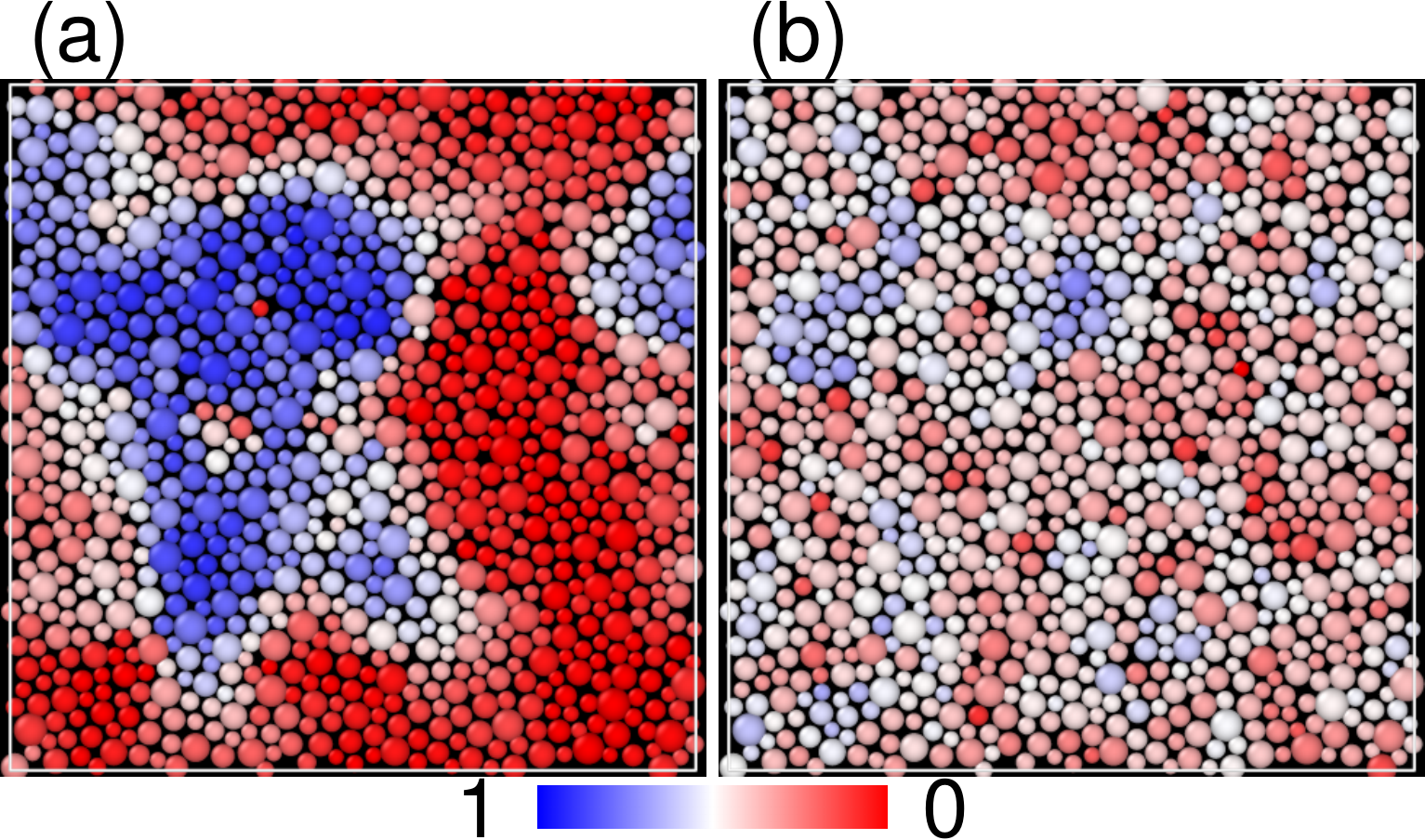}
\caption{Isoconfigurational relaxation maps in (a) MC and (b) SMC dynamics at $T=0.130$ starting from the same initial condition. Both panels show configurations at $\tau_\alpha = \num{2.0e7}$ in (a) and $\tau_\alpha = \num{4.5e3}$ in (b).}
\label{fig:snapshots_iso}
\end{figure}

To get a qualitative feeling for the degree of correlation between MC and SMC isoconfigurational dynamics, we show in Fig.~\ref{fig:snapshots_iso} an example of a map of isoconfigurationally-averaged local relaxation functions for the same initial condition but using the two dynamics. We choose the temperature $T=0.130$ in 2D, where the MC dynamics is already glassy ($\tau_\alpha=\num{2.0e7}$), but SMC dynamics is rather fast ($\tau_\alpha=\num{4.5e3}$). Both maps are taken at the respective $\tau_\alpha$. The MC map in Fig.~\ref{fig:snapshots_iso}(a) displays a clear spatial separation between relaxed (red) regions and unrelaxed (blue) ones, thus revealing spatially extended dynamic correlations and a large dynamic correlation lengthscale. For this particular system, this heterogeneous pattern is known to be largely controlled by dynamic facilitation~\cite{Scalliet2022Thirty,Herrero2024PRL}, but it also correlates with various structural quantities~\cite{Widmer_Cooper_2008,Schoenholz_2016,Tong_2018,Lerbinger2022}.

The SMC map in Fig.~\ref{fig:snapshots_iso}(b) is strikingly different. The dynamic contrast between relaxed and unrelaxed regions is suppressed while the spatial extension of dynamic correlations is also much smaller. These observations offer microscopic support to the suppression of single-particle heterogeneity discussed in Sec.~\ref{sec:single_DH} and of collective relaxations in Sec.~\ref{sec:collective}. 

Looking more carefully in the SMC map, we do observe some spatial heterogeneity revealing for instance bluer regions that are more stable than average. For instance the top left corner has more blue particles, which seems reminiscent of the large blue region which appears in the MC map. This may indicate that structurally stable regions influence both dynamics, but give rise to distinct dynamic heterogeneity patterns. 

\subsection{Correlation between the two dynamics}

To quantify the similarity between MC and SMC isoconfigurational relaxation maps, we introduce the following Pearson correlation coefficient
\begin{align}
\rho_P(
\bm{x}(t_\mathrm{MC}),
\bm{y}(t_\mathrm{SMC})
)
=
\frac{\sum_{i} \delta x_i \delta y_i }{\sqrt{\sum_{i}  \delta x_i^2} \sqrt{\sum_{i}  \delta y_i^2}},
\end{align}
between the vector $\bm{x} = (f_s^1(q,t_\mathrm{MC}), \dots, f_s^N(q,t_\mathrm{MC}))$ of local relaxation in the MC isoconfigurational ensemble, and that of the SMC ensemble, $\bm{y} = (f_s^1(q,t_\mathrm{SMC}), \dots, f_s^N(q,t_\mathrm{SMC}))$, with $\delta x = x - \braket{x}$ the deviation from the average. In the calculation of the Pearson coefficient, we average over five independent initial configurations.

\begin{figure}
\includegraphics[width=\linewidth]{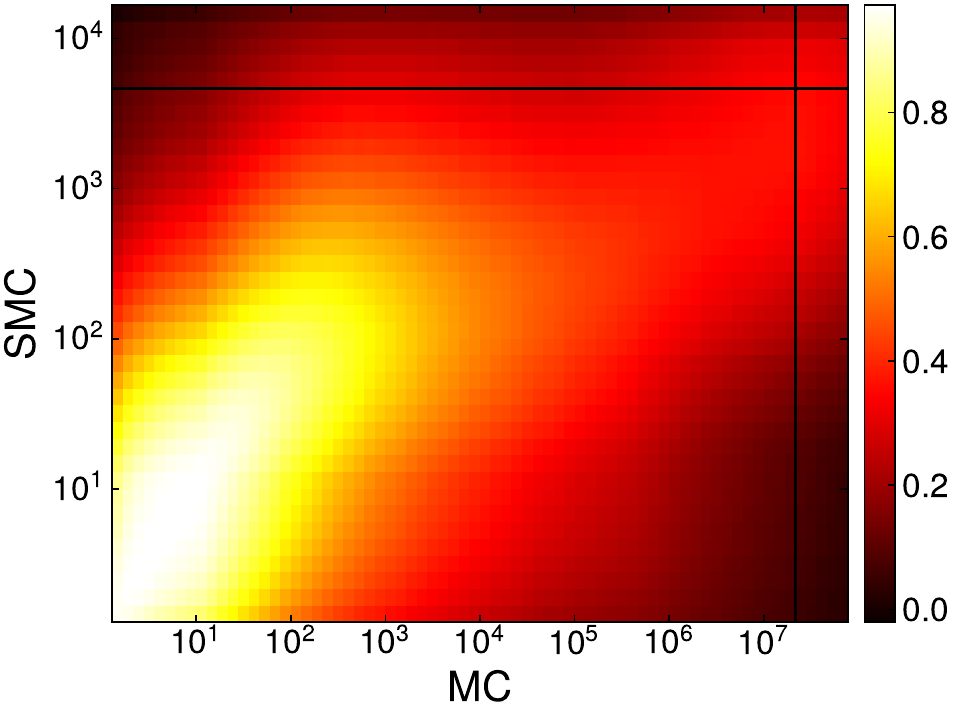}
\caption{Pearson correlation coefficient between MC dynamics at time $t_\mathrm{MC}$ and SMC dynamics at time $t_\mathrm{SMC}$. The black lines correspond to the structural relaxation time $\tau_\alpha$ in MC and SMC dynamics, respectively.}
\label{fig:pearson}
\end{figure}

Since the Pearson coefficient depends on the times chosen in each dynamics to quantify the local dynamics, we present in Fig.~\ref{fig:pearson} a heat map of its evolution with both quantities $t_\mathrm{MC}$ and $t_\mathrm{SMC}$, where the relaxation times $\tau_\alpha$ are indicated with black lines. The correlation between both dynamics is maximal at very short times with a large Pearson correlation, $\rho_P > 0.97$. This is reasonable since at short times, swap moves have very little influence on the dynamics.  However, the correlation degrades rapidly as times grow towards the structural relaxation times, to reach about $\rho_P = 0.4$ at $\tau_\alpha$.
The modest degree of correlation rationalises the above observations regarding the important differences in various signatures of glassy dynamics. We also note that although modest, the correlation is not vanishing (the Pearson coefficient is zero for uncorrelated variables), suggesting the possibility that some underlying structural features can indeed be predictors of both dynamics. It would be useful to apply for instance machine learning tools used to predict the MC dynamics to also predict the SMC one.   

\section{Discussion}

\label{sec:discussion}

We have performed extensive simulations of two glass-forming liquids to offer a detailed comparison between SMC dynamics with conventional MC. In addition to the known acceleration of the bulk dynamics, we have discovered a strong suppression of dynamic heterogeneity, with a smaller contrast between fast and slow particles, together with a drastic shortening of spatial correlations of the dynamics. We also showed that even when comparing MC and SMC on a more equal footing using plots parameterized by the growing relaxation time, the strong suppression of dynamic heterogeneity persists. 

A central numerical result is the relation between the growth of spatial correlations of the dynamics, which is different from observations using conventional MC. However, the decoupling between dynamic and static fluctuations observed for MC is also very much suppressed in SMC, and the behaviour of dynamic correlations in SMC is very similar to measurements of static correlations. The growth of the point-to-set correlation lengthscale is similar to the one of the SMC dynamic correlation lengthscale, the Adam-Gibbs relation is satisfied with reasonable values of the dynamic exponent $\psi$.  

Let us now turn to a physical interpretation of these results. We first recall that although it is non-physical, in the sense that swap moves do not appear during the conventional MC or MD dynamics, the SMC dynamics can nevertheless be seen as a local dynamics, in which the diameters are now fluctuating quantities~\cite{Ninarello_2017}. As such, it represents a valid microscopic dynamics that can in principle be studied using theoretical tools similar to the MD dynamics. It is thus a totally valid question to ask about its physical interpretation and relation to available theoretical approaches of the glass transition dynamics.    

It was proposed, in particular, that SMC dynamics corresponds to an ordinary glassy dynamics, but with a mode-coupling crossover temperature that it is depressed to lower temperatures~\cite{Ikeda_Zamponi_Ikeda_2017,Szamel_2018,Brito_2018}. We find no strong evidence from the numerical simulations for this interpretation, beyond the obvious fact that the temperature dependence of the bulk relaxation time is indeed shifted to lower temperatures, as in Fig.~\ref{fig:tau_alpha}. The observation of suppressed deviations from non-Gaussian dynamic behaviour coupled to a growth of the dynamic susceptibility maybe reminiscent of the reported behaviour found in the Gaussian core model~\cite{Ikeda_PRL_2011,Coslovich_2016}, which was indeed interpreted as a model behaving more closely according to the predictions of the mode-coupling theory of the glass transition. This interpretation is difficult to reconcile with the lack of a clear power-law description of the growth of $\tau_\alpha$ within SMC, and the strong correlation with static correlations which suggests dynamics should be better described by invoking thermally activated processes.  

A more logical interpretation of the numerical results is that the swap Monte Carlo moves, by introducing stronger fluctuations at the local scale (leading for instance to enhanced vibrational motion at short times~\cite{Ninarello_2017}), effectively reduce the influence of local kinetic constraints that contribute to the slowdown of the physical MC dynamics. As a result, the SMC dynamic heterogeneity becomes less pronounced. The observation that dynamic correlations are spatially less extended than in MC also suggests that dynamic facilitation must be much less relevant for SMC dynamics, a suggestion that should be confirmed in future work using for instance the approach of Refs.~\cite{Scalliet2022Thirty,Herrero2024PRL}. This interpretation is also in line with a recent work emphasizing the role of local kinetic constraints on glassy dynamics~\cite{Gavazzoni2024Testing}. Our conclusions are also consistent with the ones obtained from the study of a kinetically constrained model~\cite{Gutirrez2019}.  

For the two polydisperse models considered here, where swap Monte Carlo is extremely efficient, the dynamic correlation lengthscale of the SMC dynamics seems very close to the point-to-set lengthscale, while the Adam-Gibbs relation is obeyed. Our interpretation is that the suppression of local barriers is so efficient in these two models, that the activated dynamics involving collective rearrangements of particles over a lengthscale given by the point-to-set lengthscale is at play, following the finite dimensional predictions of the RFOT theory. A logical corollary is that RFOT theory does not describe well the physical MC dynamics~\cite{Wyart_2017,Gavazzoni2024Testing}. In future work, one should also test this interpretation, for instance by comparing the dynamic correlations revealed by the SMC dynamics to the spatial fluctuations of the static Franz-Parisi potential measured in Ref.~\cite{Guiselin_selfinduced_2022}, or the local fluctuations of the configurational entropy~\cite{Berthier_2021}.

The rigorous relations between timescales and lengthscales discussed in Ref.~\cite{Montanari_2006}, together with the correlation between SMC dynamics and static fluctuations suggest that for these polydisperse models, local barriers are so efficiently suppressed that the slowdown of the SMC dynamics at low temperature is mostly controlled by the growth of the point-to-set lengthscale, and thus ultimately, by the approach to a random first-order transition, akin to a Kauzmann transition at temperature $T_K$. This would for instance explain the extreme acceleration of the dynamics in 2D where $T_K \approx 0$~\cite{Berthier_2D_2019}. A direct consequence of this interpretation, however, is that for these models, no local Monte Carlo algorithm will be able to improve much the performances of the SMC algorithm. This suggests that future algorithms to equilibrate supercooled liquids even faster should be able to perform collective Monte Carlo that can effectively lower the value of the dynamic exponent $\psi$. For this, alternatives to the SMC algorithm should be developed that involve more collective moves~\cite{Dress_1995,Michel2014,PhysRevLett.133.028202,berthier2024monte} or non-equilibrium approaches~\cite{PhysRevLett.131.257101}.   

\begin{acknowledgments}
We thank Misaki Ozawa, Cecilia Herrero, Camille Scalliet, Hayato Shiba, Matthieu Wyart for sharing their data and for valuable discussions. This work is supported by the Japan Society for the Promotion of Science (JSPS) through the Overseas Research Fellowships (K. S.).
\end{acknowledgments}

\end{document}